\documentclass[10pt]{iopart}
\maketitle
\usepackage{graphicx}

\usepackage{iopams}  
\begin{document}

\title[Thermally-induced nonlinear spatial shaping in nematic liquid crystals]{Thermally-induced nonlinear spatial shaping of infrared femtosecond pulses in nematic liquid crystals}

\author{Vittorio Maria di Pietro$^{1,2}$,Aur\'elie Jullien$^1$, Umberto Bortolozzo$^1$, Nicolas Forget$^2$, Stefania Residori$^1$}

\address{$^1$ Institut de Physique de Nice, Universit\'e de Nice Sophia-Antipolis, CNRS UMR 7010, 1361 route des Lucioles, 06560 Valbonne, France.\\}
\address{$^2$ Fastlite, Les collines de Sophia, 1900 route des cretes, 06560 Valbonne, France. \\}
	
\ead{vittorio.dipietro@fastlite.com}
\vspace{10pt}
\begin{indented}
\item[]August 2017
\end{indented}

\begin{abstract}
An optically-induced thermal non-linear effect in a nematic liquid-crystal (E7) cell is evidenced through weak light-absorption by the ITO coating of an infrared pulsed femtosecond laser ($\lambda_0=1.5 \mu m$). Strong spatial self-phase modulation is generated, thus a multiple-ring pattern is observed in the far-field. The sign of the nonlinearity is changed depending on the laser polarization. The refractive index and thermal gradients are measured as a function of the laser intensity and we observe that the temperature can increase close to the nematic phase transition. The fidelity and stability of the process open new prospects for spatial shaping devices and delimits the operating wavelength range for ultrafast liquid-crystal based electro-optic application. 
\end{abstract}


\newpage

\section{Introduction}
Spatial self phase modulation is known to induce multiple-ring pattern when a laser beam propagates through a non-linear medium, thin enough to prevent significant contribution from self-focusing or defocusing \cite{Garcia2010}. As a highly nonlinear medium \cite{Khoo2009}, thin layers of liquid crystals (LC) are particularly suited to induce such effects. Actually, multiple-ring pattern and self diffraction in a two-wave configuration are reported in numerous studies \cite{Khoo1987, Khoo2014, Sanchez, DeLuca}. The original process can be instantaneous Kerr-effect, but also field-induced molecular orientation generating refractive index changes. Self-phase modulation rings (up to several tens of rings) have been observed with high-power continuous laser light in homeotropic cells \cite{Durbin1981} or dye-doped LC to enhance the nonlinearity \cite{Ara2009, Ono1998}. 
Furthermore, transient self-diffraction can also be induced by thermal changes in the LC medium, mostly due to light absorption in a thin dye-doped LC, with a temperature increase of a few degrees \cite{Khoo1985, Ono2000}. Heating of the electrode coating, so as to generate a transient nonlinear grating in a thin cell, has also been reported \cite{Kuzhelev2003}. Most of these studies have been realized with continuous or nanosecond laser sources. Driving nonlinear modification of the optical index in nematics with ultrashort laser pulses remains occasional so far. Pulsed peak intensity above the $GWcm^{-2}$ has recently lead to the measurement of coherently excited Kerr effect \cite{Cattaneo}. Self-action effect of 100 fs pulses in a thin absorbing nematic layer has also been reported \cite{Gayvoronski}. 

In addition to being efficient nonlinear media, LC offer numerous prospects in photonics. For instance, nematics LC-based electro-optical devices enables tunable phase-shifting of optical and THz radiations\cite{Yang2014, Wu1984, Jullien2016}, or phase-sensing in a light-valve configuration \cite{Bort2013}. Indium-Tin-Oxide (ITO) is one of the most employed electrode coating for electro-optic applications, thanks to its good transmission in the visible part of the optical spectrum. However, extension of these devices to the infrared light rises together interest and difficulties, as ITO presents a strong absorption and nonlinear behavior around 1.55 $\mu$m \cite{Boyd}. Although some optimized versions of ITO are available in this spectral range, a remaining weak absorption might have a significant impact on the LC optical answer, as suggested in a recent theoretical study \cite{He2017}. 

In this letter, we demonstrate that an infrared femtosecond oscillator undergoes strong spatial self-phase modulation in a thick E7 nematic mixture, due to partial laser light absorption ($\sim 20\%$) in a yet optimized ITO coating. The multiple-ring pattern analysis enables us to study the induced thermal nonlinearity for both polarization directions. This experiment enables to control in a reversible way both refractive indices up to the isotropic phase transition, that is a thermal gradient as high as $40K$ across the Gaussian laser spot. 
A power density threshold for electro-optical LC cells in the near-infrared is then set up. 
Furthermore, such a strong nonlinearity with reversible sign depending on the polarization and  without beam collapse due to self-focusing represents a novel insight for femtosecond optical pulses. 
Finally, this effect is characterized by the confinement of the thermal gradient to the laser spot and long-term stability. Thus, numerous applications can be contemplated, among them spatial shaping \cite{Cheng2017}, beam measurement and spatio-temporal shaping of femtosecond pulses. 

\section{Experimental setup}

\begin{figure}[h!]
	\centering
	\includegraphics[width=10cm,height=30cm,keepaspectratio]{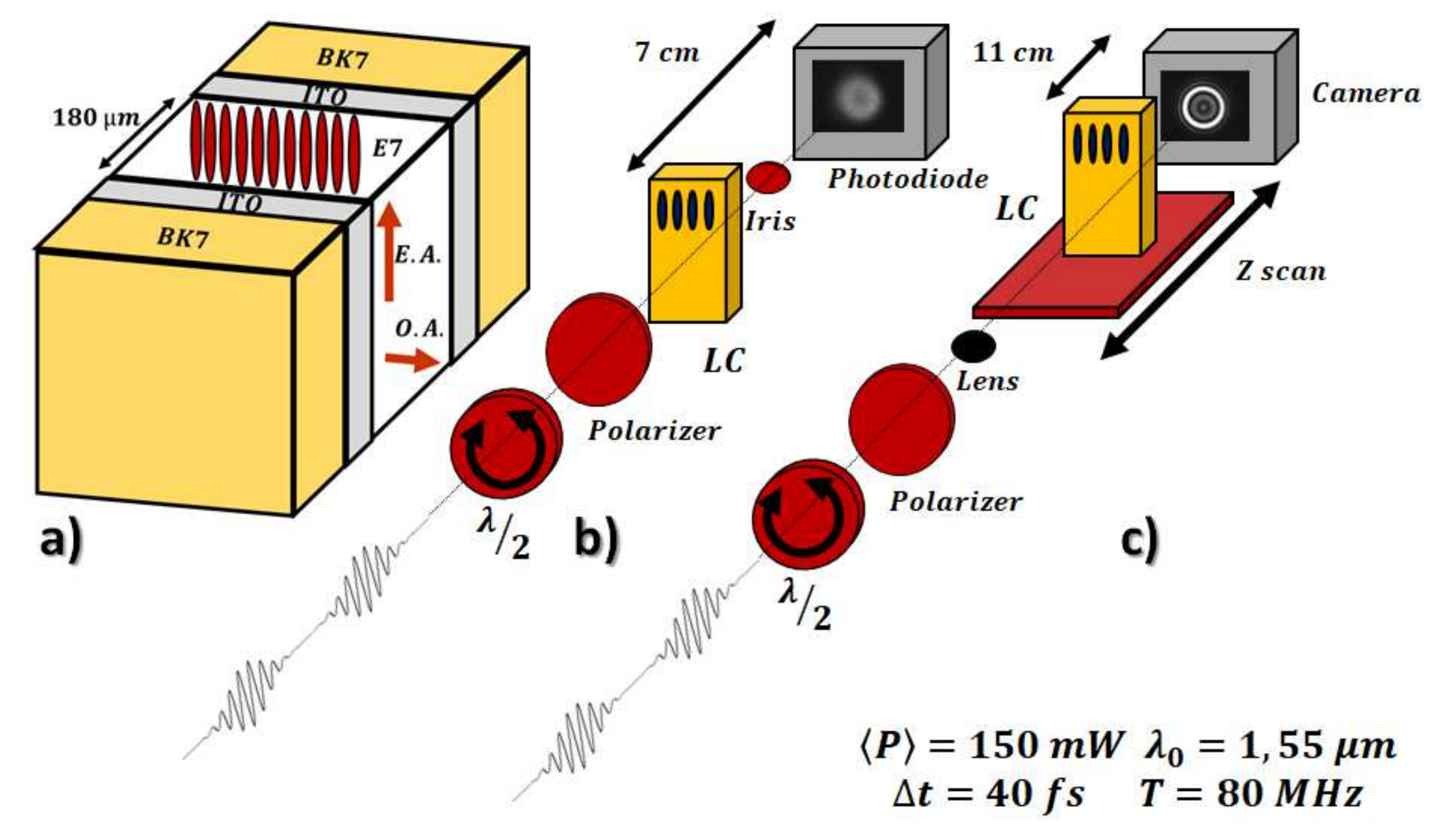}	
	\caption{\small (a) Sketch of the LC-cell. The LC anchoring direction defines the molecular director and the extraordinary axis (E.A.) of the birefringent medium. The ordinary axis (O.A.) is also indicated. (b,c) Experimental setups. A half-wave plate and a polarizer set the laser polarization and pulse energy.I n b), the laser  is collimated and an iris selects the central part of the beam after propagation in the LC cell. In c), the laser is focused onto the LC cell and the beam profile is acquired after propagation (Goldeye ALLIED vision P-008 SWIR). }
	\label{expsetup}
\end{figure}

The main LC cell used in the experiment is schemed in {\bfseries\figurename~\ref{expsetup}a}. This cell is home-made with conductive windows from Thorlabs. These windows are composed of a 5 mm thick AR-coated BK7 substrate on the top of which is deposited a thin layer of ITO optimized for the spectral range 1050-1650 nm. A thin film of polyvinyl alcohol (PVA) is spin-coated on the ITO layer and then rubbed with subsequent strong anchoring to align the molecules in a plane parallel to the substrate (planar configuration). Finally, the nematic mixture E7 is inserted between them with a  layer thickness of 180$\mu$m. No voltage is applied for this experiment. The LC-cell transmission is then 75$\%$ in the considered spectral range, consistent with the window transmission (89$\%$ per window, as provided by the furnisher).
The employed femtosecond laser (TOPTICA) produces a 40 fs pulses train, with a high repetition rate of 80 MHz and a spectral bandwidth centered at  $\lambda_0=1.55 \mu m$ (1.3 $\mu$m-1.7 $\mu$m). The output average power is 150 mW. The beam profile is Gaussian (2 mm diameter).  The maximum peak intensity and average power density are indicated in  {\bfseries\tablename~\ref{table}}. 

\begin{table}[h!]
	\caption{\label{table} Maximum peak intensity and average power density in both experimental set up configurations.}
	\begin{indented}
		\item[]\begin{tabular}{lclclcl}
			\br
			Configuration&Peak Intensity&Average Power Density\\
			\mr
			 b)&$1.5$  $ MW/cm^{2}$&$5$  $ W/cm^{2}$\\
			c)&$2.5$  $ GW/cm^{2}$&$7.6 $  $ kW/cm^{2}$\\
			\br
		\end{tabular}
	\end{indented}
\end{table}

\begin{figure}[h]
	\centering
	\includegraphics[width=10cm,height=30cm,keepaspectratio]{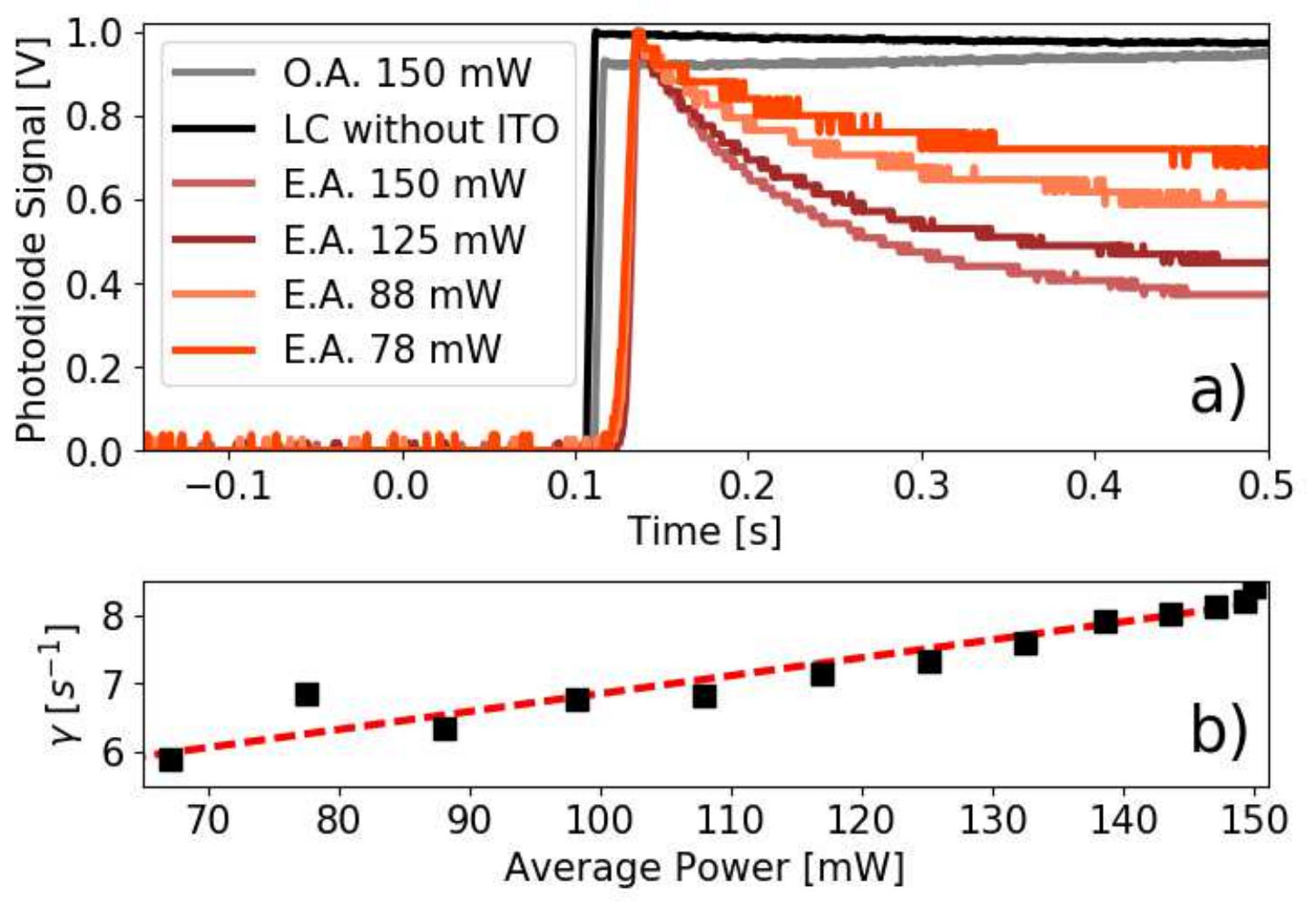}	
	\caption{\small a) Signal transmitted through a central iris (Fig.~\ref{expsetup}b) in different configuration (O.A. = ordinary axis, E.A. = extraordinary axis, see text).  b) Characteristic decay rate $\gamma$ $(s^{-1})$ of an exponential fit of the signal decrease, as a function of laser average power. The dashed line indicates a linear fit.}
	\label{time} 
\end{figure}

The first evidence of the thermal effect simply occurs with the collimated beam, as illustrated in {\bfseries\figurename~\ref{expsetup}b}. The laser beam, after passing through the LC-cell whose extraordinary axis is parallel to the laser polarization direction, exhibits  a spatial deformation after about 10 cm of propagation. A power depletion in the center of the beam becomes visible, depending on the polarization direction and average power. An iris is then used to select the central part of the beam and the transmission is measured for various configurations, in order to establish the temporal dynamic of the process. The results are shown in {\bfseries\figurename~\ref{time}a}. 
At first, we underline that the same measurement performed on another LC-cell with the same thickness but no ITO layer shows a constant transmitted signal in time whatever the laser polarization. This proves the significant role of the electrode on the process and rules out pure laser-induced molecular reorientation. 
Then, when the pulses polarized along the extraordinary direction (E.A.) propagate through the ITO-coated LC-cell, center beam depletion occurs and the transmitted signal decreases in a few hundreds of milliseconds. The non instantaneous character of the process suggests a thermal effect. 
To confirm this, the laser pulse duration is changed by adding more than $\pm 1500 fs^2$ (i.e. roughly a factor of 4 on duration) and no noteworthy variation with respect to the original signal is detected. 
This can be considered as a proof that the generated effect depends only on the laser power density rather than the peak intensity. The remaining explanation resides in light absorption in both ITO layers creating a thermal index gradient in the LC cell. 
Increasing the laser power fastens the exponential decay of the transmission, with a linear dependence, as shown in  {{\bfseries\figurename~\ref{time}}} and as an additional indication of the absorption-induced thermal nature of the effect.
When the laser beam is polarized along the ordinary axis (O.A.), almost no change of the transmission is detected, establishing  the current experimental conditions (e.g. $ 5 {W}\!/{cm^2}$) as the threshold of the studied thermal effect ({\bfseries\tablename~\ref{table}}).  

\section{Experimental results and analysis}

In order to increase the average power density, the infrared beam is focused onto the cell ($f = 40mm$), positioned on a translation stage ({\bfseries\figurename~\ref{expsetup}c}). Rotating the half-wave plate and moving the cell with respect to the focus enables to control the power density, with a maximum value of $7.6 kW/cm^2$ ({\bfseries\tablename~\ref{table}}). Images of the laser spot are acquired after $11 cm$ of propagation. 
As expected, the nonlinear effect is significantly stronger in this configuration and the generation of a multiple-ring pattern is observed. This pattern is typical of spatial self-phase modulation in thin media, when the dephasing across the Gaussian beam profile is larger than $2 \pi$ {\cite{Garcia2010}}. 
Spatial interference profiles are created and the number of rings ($N$) is a function of the index gradient, here thermally-induced, and, therefore, increases with the power density. We propose the following explanation. Heat absorption in the ITO is transfered to the nematic layer, whose heat transfer coefficient ($h$) is two orders of magnitude lower than the electrode and one order of magnitude lower than the glass ($h_{E7}=0.13 W m^{-1} K^{-1}, h_{ITO}=11 W m^{-1} K^{-1}, h_{BK7}=1.11 W m^{-1} K^{-1})$\cite{He2017}. The heat then remains confined in the LC, enabling the establishment of a thermal gradient. Large and nonlinear sensitivity of the thermotropic LC with the temperature generates the observed refractive index changes \cite{Khoo2014b}.

\begin{figure}[h]
	\centering
	\includegraphics[width=\columnwidth]{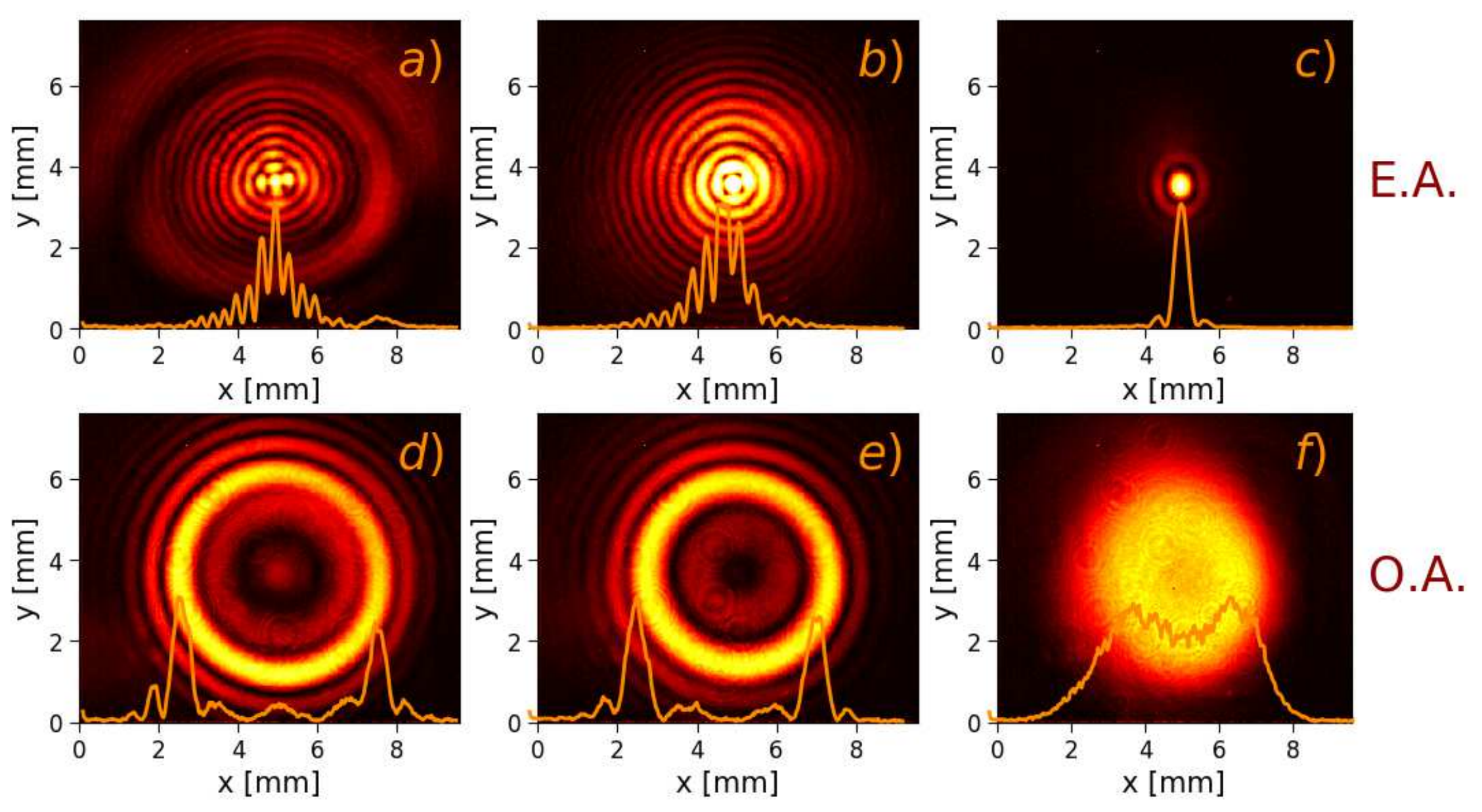}	
	\caption{\small Ring pattern formation. First line : the laser is polarized along E. A. (a=  $150 mW $, b=  $ 132 mW $, c= $108 mW $ ). Second line: the laser is polarized along O. A. (d=  $150 mW $, e=  $143 mW $, f= $108 mW $ ).}
	\label{ring}
\end{figure}

\begin{figure}[h]
	\centering
	\includegraphics[width=11cm,height=30cm,keepaspectratio]{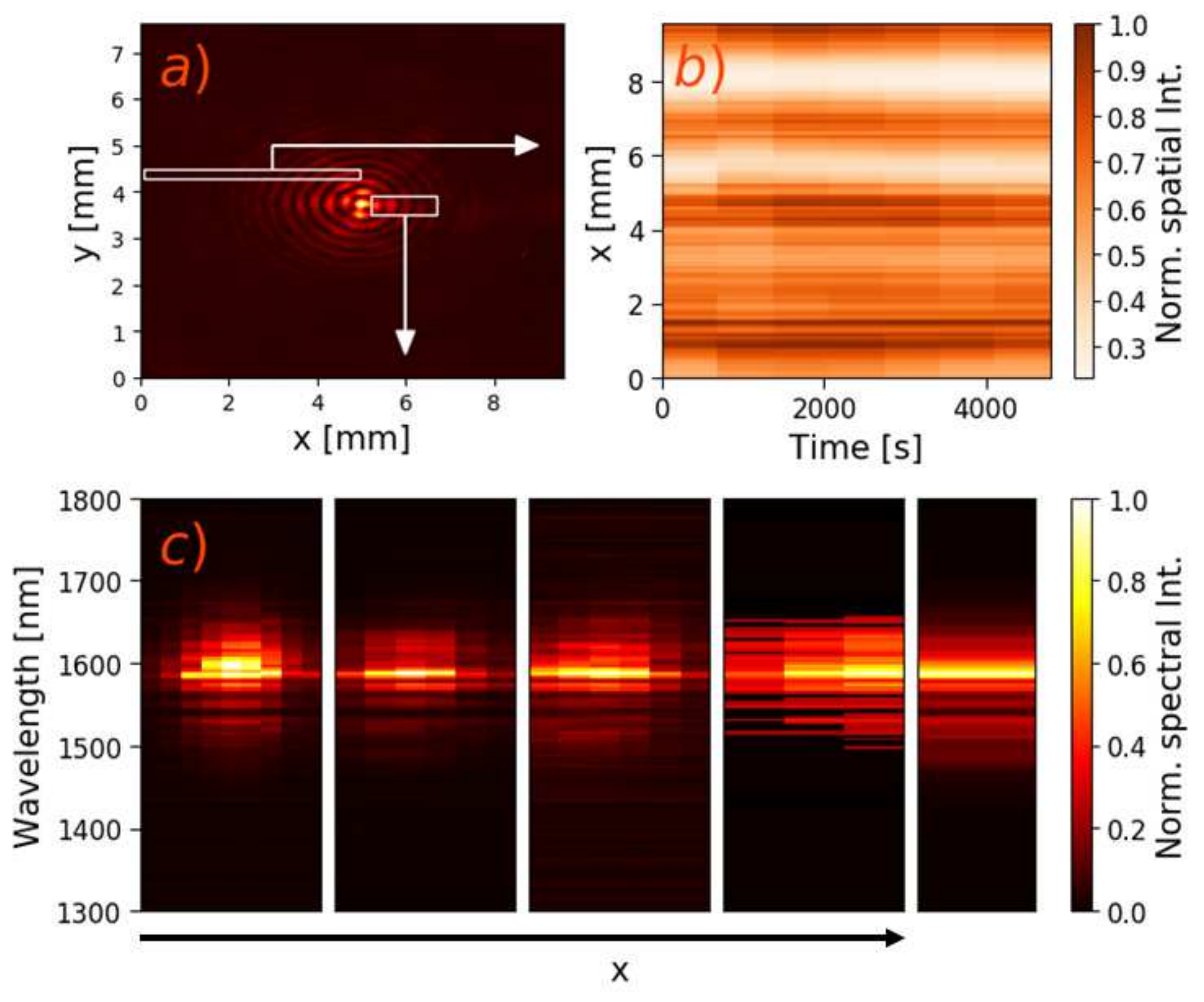}	
	\caption{\small 
		a)  Acquired multiple-ring pattern (laser polarized along EA). b) Horizontal spatial distribution as a function of time for $N=15$. c) Horizontal spectral distribution. Each sub-plot is related to one-ring. The righter plot is the reference laser spectrum.}
	
	\label{stab}
\end{figure}

Some typical images are shown in {\bfseries\figurename~\ref{ring}} for both extraordinary ($a,b,c$) and ordinary ($d,e,f$) laser polarization directions. The LC-cell is out of focus and the laser power is changed. As expected from the thermotropic character of the LC birefringence, the sign of the nonlinearity is opposite, according to the laser polarization. If the laser is polarized along the cell extraordinary axis, a central hot spot appears and  the number of rings increases with the energy. In a given observation plane, the ring size is found constant (250 $\mu$m in this case) \cite{Durbin1981}. Conversely, if the laser is polarized along the ordinary axis, some large rings surround a central beam depletion. Increasing the intensity, the black spot remains, but the first ring intensity increases and, at the same time, more rings appear. This different behavior is in agreement with simulations of \cite{Garcia2010} about the ring formation on a thin nonlinear medium. 
A second observation resides in the large rings number measured for the extraordinary beam, indicating a strong and confined thermal gradient. We performed some stability measurements for a given rings pattern ($N$={15}) and we found that, over more than one hour, the number and spatial position of the rings keep constant ({{\bfseries\figurename~\ref{stab}a}}). No variation of the spatial dephasing is measured within the resolution of the camera (30 $\mu$m), e.g. the stability is estimated better than {$ {\pi}\!/{5}$ for a maximum phase shift of $30 \pi$} . Therefore the refractive index variation is temporally stable and the thermal gradient is well-confined.  This highlights the exceptional thermal properties of LC.

The presented effect is original with respect to the usual spatial nonlinear effects achieved with femtosecond pulses. Indeed, bulk media are often preferred. The longer non linear length thus generates strong self-focusing or defocusing, preventing to acquire such high spatial phase shift without beam collapse. Furthermore, the process being due to instantaneous Kerr effect, temporal self-phase modulation occurs as well. Here, no modulation of the temporal phase of the ultrafast pulse occurs, given the timescale of the involved phenomenon, and thus the spectro-temporal characteristics of the pulse are not affected. As a matter of fact, the spectral distribution in the rings is found unchanged with respect to the original spectrum ({{\bfseries\figurename~\ref{stab}b}}).

\begin{figure}[h]
	\centering
	\includegraphics[width=\columnwidth]{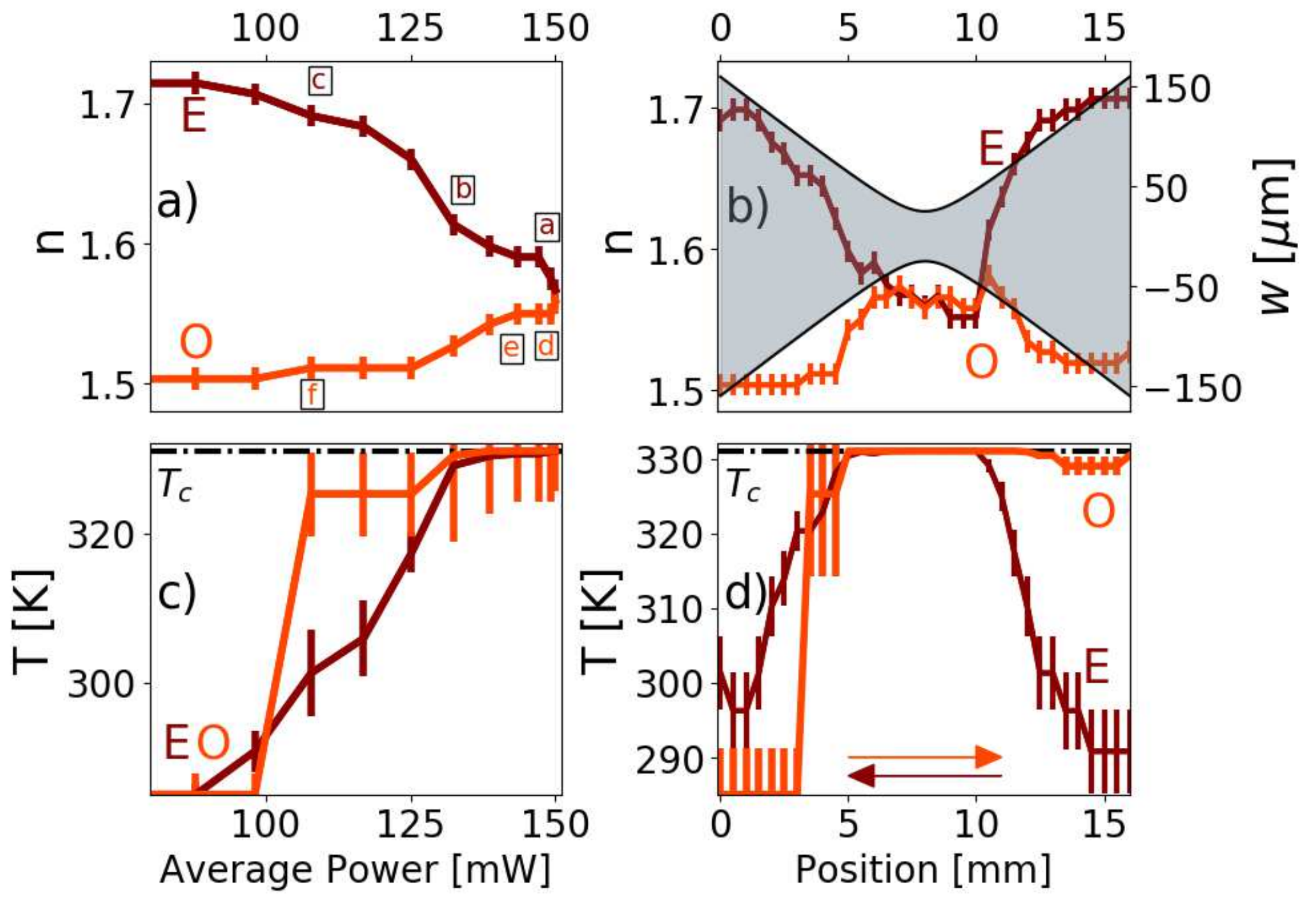}	
	\caption{\small Refractive index $n$ variation as a function of the average power and the position around the focus (a, b respectively). The red (resp. orange) line refers to an extraordinary (resp. ordinary) laser polarization direction. In a), the labels refer to {\bfseries\figurename~\ref{ring}}. 
		The beam waist evolution appears in gray in panel b).	
		In c), d) are plotted the maximum LC temperature $T$ as function of the average power (c) and the position (d), deduced from a) and b) (see text). Again, the red (resp. orange) line refers to an extraordinary (resp. ordinary) laser polarization direction. Clearing temperature ($T_c=331 K$) is indicated.  Arrows show the position scan direction.}
	\label{scan} 
\end{figure}

To go further, from the scheme of {\bfseries\figurename~\ref{expsetup}c}, scans in power and position are performed in order to measure the index changes and the thermal elevation as a function of the power density.  
From the number of rings we can estimate the introduced dephasing $\Delta\varphi$, following \cite{Durbin1981}, where the following assumption was found valid :

\begin{equation}
N \simeq \frac{\Delta\varphi}{2\pi}= \frac{\Delta n L }{\lambda}.
\label{N}
\fl
\end{equation}
with $\Delta n$ the refractive index variation and $L$ the nematic layer thickness.
{\bfseries\figurename~\ref{scan}} represents the measured refractive index variation as a function of the average power (the cell is slightly out-focus) ($a$) and as a function of the position with respect to the focus, for the maximum power ($b$). The beam waist evolution is plotted as well in panel $b)$. The error bars come from an uncertainty of $ \pm  1$ ring for each measurement. The two cases of a laser beam polarized along the extraordinary axis (refractive index $n_{e}$) and along the ordinary axis (refractive index $n_{o}$) are considered.
The average power excursion shows a constant and nonlinear decrease (resp. increase) of $n_{e}$ (resp. $n_{o}$), as expected from the LC index variation with the temperature \cite{Khoo}. 
The position scan confirms this trend until extraordinary and ordinary axis are no longer distinguishable:  around the focus, the laser power density is high enough to approach the phase transition between nematic and isotropic phases (clearing point for E7: $T_c=331 K$, \cite{Li2005}) . Some background fluctuations then appear on the multiple-ring pattern, indicating the vicinity of the transition. 

Following the extensive work of Li and coauthors, who have characterized E7 thermotropy for several spectral range including near-infrared \cite{Li2005}, one can recover $\Delta T$, the LC temperature elevation across the laser spot, from $\Delta n$.  Considering the two laser polarization directions, we then exploit the following laws:

\begin{equation}
n_{e}(T) = A - B T + \frac{2 \Delta n_{0}}{3} \left(1-\frac{T}{T_c}\right)  ^\beta
\end{equation} 
\begin{equation}
n_{o}(T) = A - B T - \frac{ \Delta n_{0}}{3} \left(1-\frac{T}{T_c} \right)^\beta
\end{equation} 
where $A=1.723 , B = 5.24$ $ 10^{-4} , T_c = 331 , \beta  = 0.2542, \Delta n_{0}=0.3768$ (constants for E7 liquid crystal mixture) \cite{Li2005}. The initial LC temperature is $285$ $K$. As a result, the thermal gradient $\Delta T$ is calculated and the maximum temperature is plotted in {\bfseries\figurename~\ref{scan} c, d}. $T_o$ (resp. $T_e$) then refers to the laser polarized along the ordinary (resp. extraordinary) axis. The strong non-linearity of the index evolution increases the error bars when approaching $T_c$. These figures highlight several physical characteristics of the described phenomenon. 
At first, in the power scan ({\bfseries\figurename~\ref{scan} c}), a difference between the extraordinary and the ordinary polarization directions is visible. For a given power density, $T_o > T_e$. This is due to the anisotropy of the heat transfer coefficient. According to \cite{Mercuri1998, Khoo2014}, this thermal coefficient is nearly two times lower in the ordinary direction, confining even more the transfered heat. It can further be noticed that, although the index variation is strongly nonlinear, the temperature $T_e$ increases linearly with the power, before reaching saturation close to the transition. For the lower thermal gradients, this linear dependence was predicted in \cite{He2017}. 
Finally, the position scan temperature evolution ({\bfseries\figurename~\ref{scan}d}) is not symmetric around the focus, because of the scan acquisition direction (opposite for both polarization directions). After getting close to the transition, the recovery time of the LC molecules is considerably increased but the initial birefringence state and molecular orientation are finally recovered after a few minutes. 

\begin{figure}[h]
	\centering
	\includegraphics[scale=0.5]{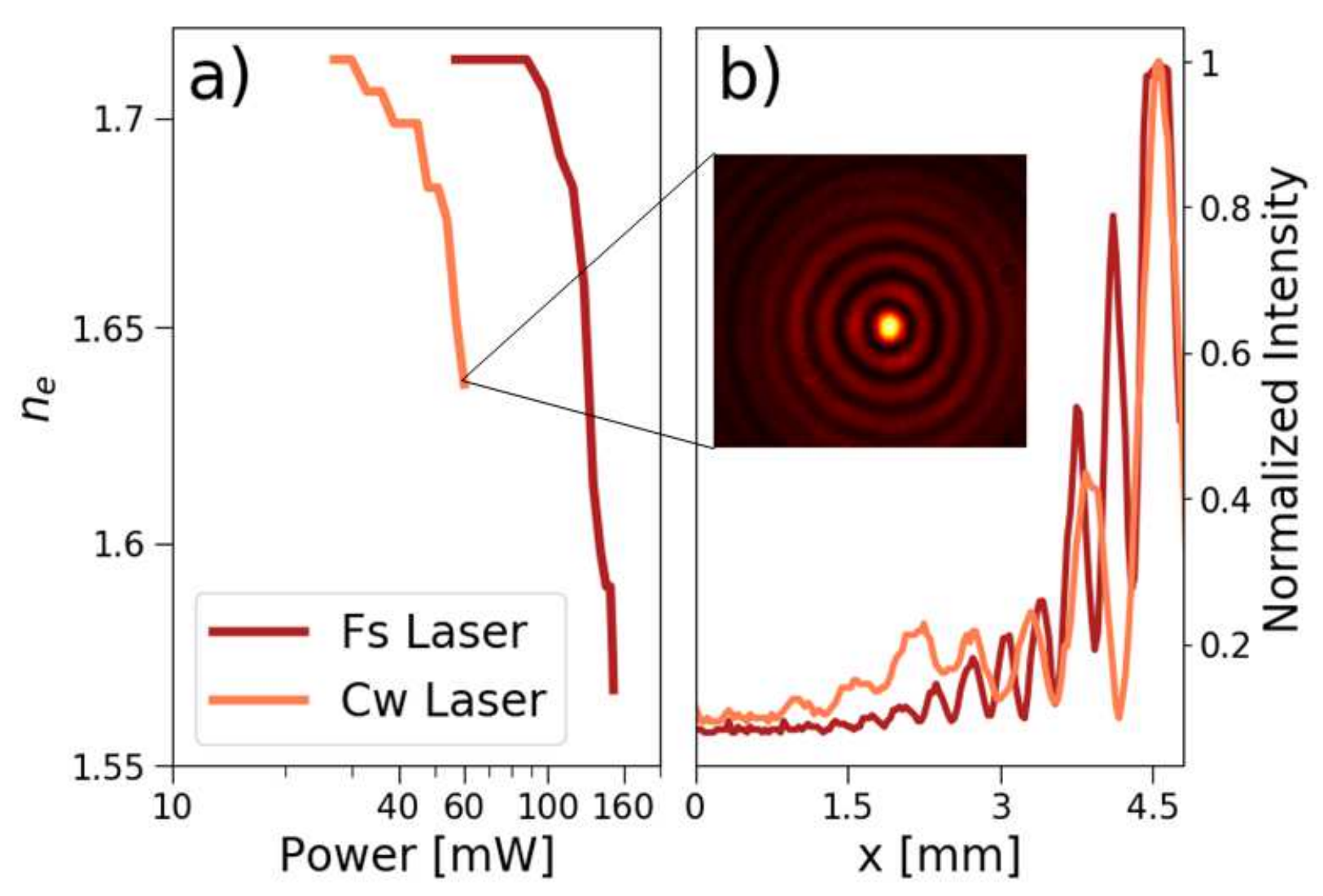}	
	\caption{\small a) Refractive index variation ($n_{e}$, the laser is polarized along E.A.) as a function of the average power for the femtosecond laser (dark red curve) and cw laser (orange curve) in a log-log scale. b) Corresponding normalized intensity horizontal profiles for  $\Delta n_{e} = 0.08$, e. g. $N\sim10$. Inset : ring pattern observed with the CW laser.}
	\label{cw} 
\end{figure}

To the best of our knowledge, this is the first time that ultrashort pulses are used to excite such a high thermal gradient in LC-cells.
We have performed the same experiment with a 100 mW femtosecond oscillator with $\lambda_0=1.06 \mu m$ and the effect was generated as well, but with a lower phase shift, following the spectral dependence of the ITO absorption \cite{Boyd}. Furthermore, the phenomenon is naturally not restricted to pulsed laser light. Similar results have been obtained with a 60 mW continuous laser at $\lambda_0=1.5\mu m$. When focused onto the LC cell, some deformation on the spot laser occurs, although the maximum phase shift is lower and  $T_c$ can not be reached. The acquired index variation  for an extraordinary polarized laser light as a function of the average power is shown in  {\bfseries\figurename~\ref{cw}a} and compared to the data acquired with the femtosecond laser, exhibiting the same dependence. The repetition rate of the pulsed radiation is therefore high enough to maintain the equilibrium of the thermal state. 
For the same number of rings (e.g. similar thermal gradient), the beam profiles are compared ({\bfseries\figurename~\ref{cw}}b). The pitch and overall shape appear to be slightly different, as in the CW case the beam profile is closer to a pure Bessel function. This is attributed to the broad spectrum of the femtosecond pulses, blurring slightly the interference pattern. 

Spatial resolution, e.g. how-well is confined the thermal gradient, has also to be considered. Numerous indications in the presented experiments tend to imply that the nonlinear effect, although thermally-induced, can be considered as localized to the laser spot. The measured stability with a focused beam, the changes observed around the 50 $\mu$m focal spot during Z-scan ({\bfseries\figurename~\ref{scan}b}), the theoretical predictions of \cite{He2017}, where the gradient does not extend beyond the finite laser dimension, all tend to foresee a spatial resolution better than 100 $\mu$m. In addition, we have performed a self-diffraction experiment in a two-wave configuration, using a 0.5W, continuous laser ($\lambda_0=1.06 \mu m$). Self-diffraction from the thermally-induced grating in the 180$\mu m$-thick LC cell has been measured for a grating period of 140$\mu m$, establishing a upper threshold for the actual spatial resolution at the optimum wavelength ($\lambda_0=1.55 \mu m$). The spatial resolution is also expected to be enhanced for a thinner LC cell. 

\section{Conclusion}

To conclude, we have demonstrated an optically-induced thermal nonlinearity in a thick nematic liquid-crystal cell, through partial light absorption ($\sim 20\%$) of the ITO coating in the infrared spectral range. We have thus been able to control in a reversible way the full dynamic of the extraordinary and ordinary indices up to the isotropic phase transition, that is a local temperature increase of $40K$. As a result, the average power density threshold is estimated to $5 {W}\!/{cm^2}$ for $\lambda_0=1.55 \mu m$. The well-confined thermal gradient enables phase-shift as high as 100 rad and significant spatial shaping of a femtosecond laser. 

The study has enabled to define the experimental conditions to achieve significant beam shaping, but also to determine the  average power density not to be exceeded to maintain the propagation of the Gaussian beam. Our observations result from the particular absorption of the employed conductive window and one can expect to balance or strengthen this effect using a different ITO layer. The absorption can indeed be fairly controlled  with more or less thick ITO layers and/or newer materials such as graphene coatings, that would present a lower absorption coefficient. 
Finally, the process can be exploited for numerous applications, among them spatial shaping and characterization of femtosecond pulses.

\section*{Funding Information}
This work is supported by the ANR (Agence Nationale de la Recherche) under the project Labcom SOFTLITE (ANR 15-LCV1-0002-01) and has received funding from the European Union's HORIZON 2020 research and innovation program under the Marie Sklodowska-Curie grant agreement No. 641272. The publication reflects only the author's view. The Research agency of the European Union is not responsible for any use that may be made of the information it contains.\\

\vspace{10pt}

\section*{List of figures}

{\bfseries\figurename~\ref{expsetup}} (a) Sketch of the LC-cell. The LC anchoring direction defines the molecular director and the extraordinary axis (E.A.) of the birefringent medium. The ordinary axis (O.A.) is also indicated. (b,c) Experimental setups. A half-wave plate and a polarizer set the laser polarization and pulse energy.I n b), the laser  is collimated and an iris selects the central part of the beam after propagation in the LC cell. In c), the laser is focused onto the LC cell and the beam profile is acquired after propagation (Goldeye ALLIED vision P-008 SWIR). \\
{\bfseries\figurename~\ref{time}}  a) Signal transmitted through a central iris (Fig.~\ref{expsetup}b) in different configuration (O.A. = ordinary axis, E.A. = extraordinary axis, see text).  b) Characteristic decay rate $\gamma$ $(s^{-1})$ of an exponential fit of the signal decrease, as a function of laser average power. The dashed line indicates a linear fit.\\
{\bfseries\figurename~\ref{ring}}  Ring pattern formation. First line : the laser is polarized along E. A. (a=  $150 mW $, b=  $ 132 mW $, c= $108 mW $ ). Second line: the laser is polarized along O. A. (d=  $150 mW $, e=  $143 mW $, f= $108 mW $ ). \\
{\bfseries\figurename~\ref{stab}}   a)  Horizontal spatial distribution as a function of time for $N=15$ (laser is polarized along EA). b) Horizontal spectral distribution. Each sub-ploy is related to one-ring. The °°° plot is the ref. laser spectrum. \\
{\bfseries\figurename~\ref{scan}} Refractive index $n$ variation as a function of the average power and the position around the focus (a, b respectively). The red (resp. orange) line refers to an extraordinary (resp. ordinary) laser polarization direction. In a), the labels refer to {\figurename~\ref{ring}}. 
The beam waist evolution appears in gray in panel b).	
In c), d) are plotted the maximum LC temperature $T$ as function of the average power (c) and the position (d), deduced from a) and b) (see text). Again, the red (resp. orange) line refers to an extraordinary (resp. ordinary) laser polarization direction. Clearing temperature ($T_c=331 K$) is indicated.  Arrows show the position scan direction.\\
{\bfseries\figurename~\ref{cw}}    a) Refractive index variation ($n_{e}$, the laser is polarized along E.A.) as a function of the average power for the femtosecond laser (dark red curve) and cw laser (orange curve) in a log-log scale. b) Corresponding normalized intensity horizontal profiles for  $\Delta n_{e} = 0.08$, e. g. $N\sim10$. Inset : ring pattern observed with the CW laser. \\

\end{document}